\documentclass[twocolumn,pre,superscriptaddress,amsmath,amssymb,longbibliography]{revtex4-1}
\usepackage{hyperref}
\usepackage{graphicx}
\usepackage{dcolumn}
\usepackage{bm}
\usepackage{float}
\usepackage{color}

\begin{document}


\title{Ultra-Efficient Reconstruction of Anisotropic Hyperuniform Continuous Random Fields in 2D and 3D via Generalized Spectral Filtering}

\author{Liyu Zhong}
\affiliation{Department of Mechanics and Engineering Science, College of Engineering, Peking University, Beijing 100871, P. R. China}
\author{Sheng Mao}
\email[correspondence sent to: ]{maosheng@pku.edu.cn}
\affiliation{Department of Mechanics and Engineering Science, College of Engineering, Peking University, Beijing 100871, P. R. China}



\date{\today}

\begin{abstract}
Hyperuniform continuous random fields suppress large‐scale fluctuations while preserving rich local disorder, making them highly attractive for next‐generation photonic, thermal and mechanical materials. However, traditional reconstruction techniques often suffer from limited spectral control or excessive computational cost, especially in high‐resolution 2D and 3D settings. In this work, we present an ultra‐efficient generative algorithm based on generalized superellipse spectral filtering, which allows independent tuning of isotropic and anisotropic spectral envelopes without resorting to costly iterative schemes. We demonstrate our method on a comprehensive set of 2D and 3D examples, showing precise manipulation of spectral band shape and orders‐of‐magnitude speedup compared to existing approaches. Furthermore, we explore the effect of simple thresholding on the generated fields, analyzing the morphological features and power‐spectrum characteristics of the resulting ±1 two‐phase maps. Our results confirm that the proposed framework not only accelerates hyperuniform field synthesis but also provides a versatile platform for systematic study of binary microstructures derived from continuous designs. This work opens new avenues for large‐scale simulation and optimized design of advanced hyperuniform materials.
\end{abstract}



\maketitle


\section{Introduction}

A wide class of engineering materials, ranging from composites \cite{zheng2019modeling} and alloys \cite{ke2019enhanced} to porous media and granular assemblies, play pivotal roles in applications from soft gripping \cite{nguyen2023liquid, elango2015review, mohammadi2021robust} to wave manipulation \cite{waveguide, damaskos1982dispersion, itin2010dispersion, kim2023extraordinary}.  These systems are characterized by richly disordered microstructures \cite{torquato2002random, Sa03a}, which complicate both the exploration of structure–property relationships and the inverse design of target functionalities.  While topology‐optimization frameworks \cite{sigmund1999design, sigmund1997design} have achieved impressive results for binary composites and continuum solids, their reliance on high‐dimensional shape parametrizations and gradient‐based solvers often leads to prohibitive computational expense.  More recently, data‐driven and materials‐informatics strategies have emerged \cite{courtright2025high, generale2024inverse, liu2024active, bessa2017framework, li2022machine, mirzaee2025inverse}, wherein one first embeds the vast microstructure space into a reduced‐dimension latent manifold and then trains analytical \cite{cang2018improving, cheng2022data} or machine‐learning models \cite{xu2022correlation} to link latent coordinates with effective properties.  Finally, optimized latent vectors are decoded back into explicit microstructures via reconstruction algorithms, a process also known as microstructure construction \cite{bostanabad2018computational, sahimi2021reconstruction, Ye98a, Ye98b}.

Among the many microstructure‐representation schemes \cite{roberts1997statistical, niezgoda2008delineation, okabe2005pore, jiao2007modeling, jiao2008modeling, hajizadeh2011multiple, tahmasebi2013cross, tahmasebi2012multiple, xu2013stochastic, xu2014descriptor, cang2017microstructure, yang2018microstructural, li2018transfer, farooq2018spectral, cheng2022data, xu2022correlation, skolnick2024quantifying, shih2024fast, casiulis2024gyromorphs}, spatial correlation functions (SCFs) and their Fourier‐space counterparts, spectral density functions (SDFs), have proven particularly powerful \cite{To02a, jeulin2021morphological, farooq2018spectral, Ch18a, shi2023computational, shi2025three}.  SCFs afford clear physical interpretation and connect rigorously to effective‐property formalisms \cite{torquato1997effective, torquato1985effective, kim2020effective, torquato2021nonlocal, kim2023effective, torquato2021diffusion, skolnick2023simulated, torquato1990rigorous, torquato2020predicting}, while SDF‐based constructions enable direct control over fluctuation suppression at selected length scales.  A popular decoder is the Yeong–Torquato (YT) approach \cite{Ye98a, Ye98b}, which formulates reconstruction as an energy minimization solved via simulated annealing \cite{kirkpatrick1983optimization}.  Although YT delivers high‐quality two‐phase microstructures with prescribed SCFs, its iterative nature and reliance on expensive forward–reverse FFT operations make it challenging to scale to large 2D and 3D domains.  In contrast, our work leverages a single‐shot, FFT‐based spectral filtering strategy that achieves ultra‐efficient reconstruction of hyperuniform continuous fields without sacrificing shape flexibility or resolution.

Disordered hyperuniform (DHU) materials combine local disorder with suppressed large‐scale fluctuations, forming a unique class of heterogeneous systems that behave like liquids or glasses at short range yet exhibit a hidden long‐range order via vanishing spectral density at zero wavenumber \cite{To03, Za09, To16a, To18a}. This interplay of disorder and order endows DHU media with remarkable properties, including anomalous wave transport \cite{ref31, ref32, scattering, granchi2022near, park2021hearing, klatt2022wave, tavakoli2022over, cheron2022wave, yu2021engineered, li2018biological}, exceptional thermal and electrical conductivity control \cite{ref34, torquato2021diffusion, maher2022characterization}, and tunable mechanical response \cite{ref35, puig2022anisotropic, kim2020multifunctional}, with potential applications across photonics, energy and multifunctional materials. Hyperuniformity has now been identified in diverse physical systems \cite{ref4, hexner2017noise, salvalaglio2020hyperuniform}, engineered composites \cite{Ge19, zhang2023approach, chen2021multihyperuniform}, and even biological tissues \cite{ref26, ge2023hidden, liu2024universal}, as reviewed in \cite{To18a}.

To date, most reconstruction efforts for DHU microstructures have focused on two‐phase media, implementing iterative decoders such as simulated‐annealing or gradient‐based schemes to match prescribed spectral densities \cite{Ch18a, shi2023computational, shi2025three}. While these methods can produce high‐quality binary composites, they suffer from substantial computational expense and limited direct control over anisotropic spectral features. Moreover, approaches for generating continuously varying property fields with hyperuniform spectra have not been thoroughly investigated. In this work, we introduce an ultra‐efficient forward generative algorithm based on generalized spectral filtering. By applying analytic superellipse-shaped masks in the Fourier domain parametrized by exponent \(p\), aspect ratios \(a,b\), and concentration parameter \(\alpha\), we directly synthesize 2D and 3D hyperuniform continuous random fields with freely tunable isotropy or anisotropy, all without iterative optimization.

We detail our implementation, which harnesses fast Fourier transforms to deliver \(O(N\log N)\) performance and minimal memory overhead for large‐scale domains. A suite of 2D examples demonstrates precise modulation of the continuous fields and their log‐power spectra under variations of \(p\), \(\alpha\), and \(a/b\). We then apply simple thresholding to produce \(\pm1\) two‐phase maps, analyzing their morphological transitions and spectral signatures. A 3D reconstruction further confirms scalability, and runtime benchmarks against conventional decoders reveal orders‐of‐magnitude speedup. This generalized spectral filtering framework thus offers a versatile, high‐performance toolkit for exploring and designing anisotropic hyperuniform structures in advanced material systems.

The remainder of this paper is organized as follows. In Sec.~II, we introduce the concept and formal definitions of continuous hyperuniform random fields, detail our ultra‐efficient reconstruction algorithm implementation (including the physical interpretation and numerical tuning of parameters \(\alpha\), \(k_{0}\), and \(\sigma\)), and describe the design of generalized superellipse spectral masks governed by \(a\), \(b\), and \(p\). In Sec.~III, we present our results: a series of 2D case studies demonstrating the influence of \(\alpha\) on continuous and thresholded ±1 field power spectra, the transition from isotropic (\(a=b\)) to anisotropic (\(a\neq b\)) morphologies, and the effect of the superellipse exponent \(p\) on spectral shape and spatial patterns, as well as a 3D reconstruction exemplifying scalability and performance. Finally, Sec.~IV offers concluding remarks summarizing the advantages of our approach, the impact of the superellipse spectral design, and outlines promising directions for future work.

\section{Definitions and Methods}
\label{definition}

\subsection{Hyperuniform, Nonhyperuniform and Antihyperuniform Random Fields}
\label{subsec:hyper}

In the context of a scalar random field, the variance of the local field over a spherical window of radius $R$ is defined as \cite{To16a,Ma17a}
\begin{equation}
\sigma_F^2(R)=\frac{1}{v_1(R)}\int_{\mathbb R^d}I(\mathbf r)\,\alpha_2(r;R)\,\mathrm d\mathbf r,
\label{eq_field_fluc_mod}
\end{equation}
where
\begin{equation}
v_1(R)=\frac{\pi^{d/2}R^d}{\Gamma\bigl(1+\tfrac{d}{2}\bigr)},
\end{equation}
and $\alpha_2(r;R)$ is the scaled intersection volume of two windows separated by distance $r$.  A random field is said to be disordered hyperuniform if its variance decays faster than the inverse window volume, i.e.,
\begin{equation}
\lim_{R\to\infty}\sigma_F^2(R)\,R^d=0,
\end{equation}
which in Fourier space is equivalently expressed by a vanishing spectral density at zero wavenumber \cite{To16a,To16b}:
\begin{equation}
\lim_{|\mathbf k|\to0} \tilde\chi(\mathbf k)=0,
\label{eq_hyper_mod}
\end{equation}
implying the autocovariance function $C(\mathbf r)$ satisfies the sum rule
\begin{equation}
\int_{\mathbb R^d}C(\mathbf r)\,\mathrm d\mathbf r=0.
\label{eq_sum_rule_mod}
\end{equation}

When the small-$k$ behavior of the spectral density follows a power-law, $\tilde\chi(\mathbf k)\sim|\mathbf k|^\alpha$ \cite{To16a}, the large-$R$ scaling of the variance falls into three hyperuniform classes \cite{To18a}:
\begin{equation}
\sigma_F^2(R)\sim
\begin{cases}
R^{-(d+1)},&\alpha>1\ \text{(Class I)},\\
R^{-(d+1)}\ln R,&\alpha=1\ \text{(Class II)},\\
R^{-(d+\alpha)},&0<\alpha<1\ \text{(Class III)}.
\end{cases}
\end{equation}
Class I systems include perfect crystals, many quasicrystals and exotic disordered networks \cite{To03,Og17,Za09,Ch18a}; Class II examples comprise certain quasicrystals, perfect glasses and maximally random jammed packings \cite{Og17,zhang2017classical,ref4,ref5,ref6,Za11c,Za11d}; Class III systems include disordered ground states, random organization models, perfect glasses and perturbed lattices \cite{Za11b,ref20,zhang2017classical,Ki18a}.

By contrast, nonhyperuniform fields exhibit weaker variance decay \cite{torquato2021diffusion}:
\begin{equation}
\sigma_F^2(R)\sim
\begin{cases}
R^{-d},&\alpha=0\ \text{(standard)},\\
R^{-(d+\alpha)},&-d<\alpha<0\ \text{(antihyperuniform)}.
\end{cases}
\label{eq_classes_nonhyper_mod}
\end{equation}
Standard nonhyperuniform examples include Poisson point processes, equilibrium hard-sphere fluids and RSA packings \cite{To18a,torquato2021local}, whereas antihyperuniform systems, such as critical fluctuations and Poisson cluster processes, possess a diverging spectral density at the origin \cite{torquato2021local}.

In the present work, we exploit this classification by tuning the exponent $\alpha$ through analytic superellipse spectral masks, enabling direct generation of continuous fields with prescribed hyperuniformity behavior across two and three dimensions.

\subsection{Spectral--Filtering Reconstruction with Generalized Superellipse Masks}
\label{subsec:spectral_filter}

We reconstruct continuous random fields by adopting the classical spectral representation technique, which dates back to the pioneering work of Shinozuka and Deodatis~\cite{shinozuka1991simulation} and Wood \& Chan~\cite{woodchan1994}.  Let $\{F_{\mathrm{wn}}(\mathbf k)\}_{\mathbf k\in\mathcal{K}}$ be a collection of i.i.d.\ complex Gaussian variables defined on the reciprocal lattice $\mathcal{K}$ of a periodic $d$--dimensional domain $\Omega\subset\mathbb R^{d}$, with $\langle F_{\mathrm{wn}}\rangle=0$ and $\langle |F_{\mathrm{wn}}|^{2}\rangle=1$.  Given a non–negative, even spectral density $\tilde{\chi}_{_\mathcal{K}}(\mathbf k)$, the filtered coefficients
\begin{equation}
F_{\mathrm{filt}}(\mathbf k)=A(\mathbf k)\,F_{\mathrm{wn}}(\mathbf k), 
\qquad
A(\mathbf k)=\sqrt{\tilde{\chi}_{_\mathcal{K}}(\mathbf k)},
\end{equation}
inherit Gaussianity and are then transformed to real space via the discrete inverse Fourier transform
\begin{equation}
f(\mathbf x)=\frac{1}{|\Omega|}\sum_{\mathbf k\in\mathcal K}
               F_{\mathrm{filt}}(\mathbf k)\,
               \exp\!\bigl(i\mathbf k\!\cdot\!\mathbf x\bigr), 
\qquad
\mathbf x\in\Omega.
\end{equation}
Because $\tilde{\chi}_{_\mathcal{K}}$ is real and symmetric, $f$ is strictly real valued; its autocovariance $C(\mathbf r)=\langle f(\mathbf x)f(\mathbf x+\mathbf r)\rangle$ and spectral density satisfy the Wiener--Khinchin pair
\begin{equation}
C(\mathbf r)=\frac{1}{(2\pi)^{d}}\int_{\mathbb R^{d}} 
                 \tilde{\chi}_{_\mathcal{K}}(\mathbf k)\,e^{i\mathbf k\cdot\mathbf r}\,d\mathbf k,
\end{equation}
\begin{equation}
\tilde{\chi}_{_\mathcal{K}}(\mathbf k)=\int_{\mathbb R^{d}} 
                 C(\mathbf r)\,e^{-i\mathbf k\cdot\mathbf r}\,d\mathbf r.
\end{equation}
Thus the choice of $\tilde{\chi}_{_\mathcal{K}}$ completely prescribes the second–order statistics of $f$ while the algorithmic cost remains $O(N\log N)$ for an $N$–point grid, owing to the FFT.

To encode both anisotropy and controlled hyperuniform scaling, we introduce the generalized superellipse norm.  In two dimensions it is defined by
\begin{equation}
K_{p,a,b}(\mathbf k)=\Bigl[(|k_{x}|/a)^{p}+(|k_{y}|/b)^{p}\Bigr]^{1/p},
\label{eq:superellipse}
\end{equation}
where \(p>0\) and the positive parameters \(a,b\) control the spectral mask’s aspect ratio.  Equation~\eqref{eq:superellipse} interpolates continuously between a diamond ($p=1$), an Euclidean circle ($p=2$), and an axis‐parallel square in the limit $p\to\infty$; moreover, $a\neq b$ renders the level sets elliptically stretched, thereby introducing a freely tunable anisotropy ratio $a/b$.  An extension to three dimensions is immediate by appending a third term $(|k_{z}|/c)^{p}$ with scale $c$.

The target spectral density is specified analytically as
\begin{equation}
\tilde{\chi}_{_\mathcal{K}}(\mathbf k)=C\,
\exp\!\Bigl[\,
  \alpha\,\ln\!\bigl(K_{p,a,b}(\mathbf k)\bigr)
  -\frac{K_{p,a,b}(\mathbf k)^{2}}{2\sigma^{2}}
\Bigr],
\label{eq:spectrum}
\end{equation}
where $C$ normalizes $\max_{\mathbf k}\tilde{\chi}_{_\mathcal{K}}(\mathbf k)=1$.  In this study we omit any offset regularization and focus exclusively on hyperuniform fields, i.e.\ $\alpha>0$.  For small wavenumbers $|\mathbf k|\ll1$, expansion of Eq.~\eqref{eq:spectrum} yields
\begin{equation}
\tilde{\chi}_{_\mathcal{K}}(\mathbf k)\sim C\,|\mathbf k|^{\alpha},
\end{equation}
so that $\alpha$ prescribes the low–$k$ power‐law scaling governing the large‐$R$ variance decay (Sec.~\ref{subsec:hyper}).  In particular, $\alpha\ge2$ corresponds to Class I hyperuniformity.

The Gaussian term with width parameter $\sigma$ confines spectral energy to an annular band centered at $|\mathbf k|=0$, with decreasing $\sigma$ sharpening this band and enhancing intermediate‐range correlations without altering the asymptotic class determined by $\alpha$.  The superellipse exponent $p$ and aspect ratio $a/b$ modulate the angular distribution of power: $p<2$ yields diamond‐like contours with pronounced diagonal lobes, whereas $p>2$ produces increasingly square‐like, axis‐aligned shells.  These spectral shaping operations translate directly into characteristic real‐space textures and influence higher‐order spatial correlations.

After inverse FFT, the raw field is demeaned and rescaled to a prescribed variance $\sigma_{f}^{2}$ via
\begin{equation}
f\leftarrow\sigma_{f}\,
            \frac{f-\langle f\rangle}{\sqrt{\mathrm{Var}[f]}}.
\end{equation}
This linear post‐processing preserves Gaussianity and the spectral form.  The overall reconstruction pipeline requires only two FFTs (one forward on white noise and one inverse on the filtered spectrum) plus minimal element‐wise operations.

Because the transformation is strictly linear, any functional map $g=f\mapsto\mathcal T[f]$ can be applied a posteriori without rerunning the spectral filter.  Of particular interest is the sign map $f\mapsto\mathrm{sgn}(f)$, which yields $\pm1$ two‐phase media whose structure factor inherits the shell‐like morphology of Eq.~\eqref{eq:spectrum}.  Although thresholding generally disturbs the exact hyperuniform scaling~\cite{ref41}, it provides a convenient probe of how the three‐parameter family $(\alpha,p,a/b)$ drives morphological transitions in binary composites.  We exploit this capability in Sec.~\ref{sec:results} to compare continuous and sign‐thresholded fields under identical spectral masks.

In summary, Eqs.~\eqref{eq:superellipse} and~\eqref{eq:spectrum} define an analytically tractable, low‐dimensional manifold of spectral densities whose independent coordinates $(\alpha,\sigma,p,a/b)$ control, respectively, hyperuniform class, radial bandwidth, angular shape, and anisotropy.  Their incorporation into a single–shot spectral filter furnishes an ultra–efficient generator for high‐resolution anisotropic hyperuniform random fields in both two and three dimensions, thereby laying the methodological foundation for the numerical and analytical studies presented in the remainder of this work.

\section{Results}
\label{sec:results}

\begin{figure*}[htbp]
  \centering
  \includegraphics[width=0.8\textwidth]{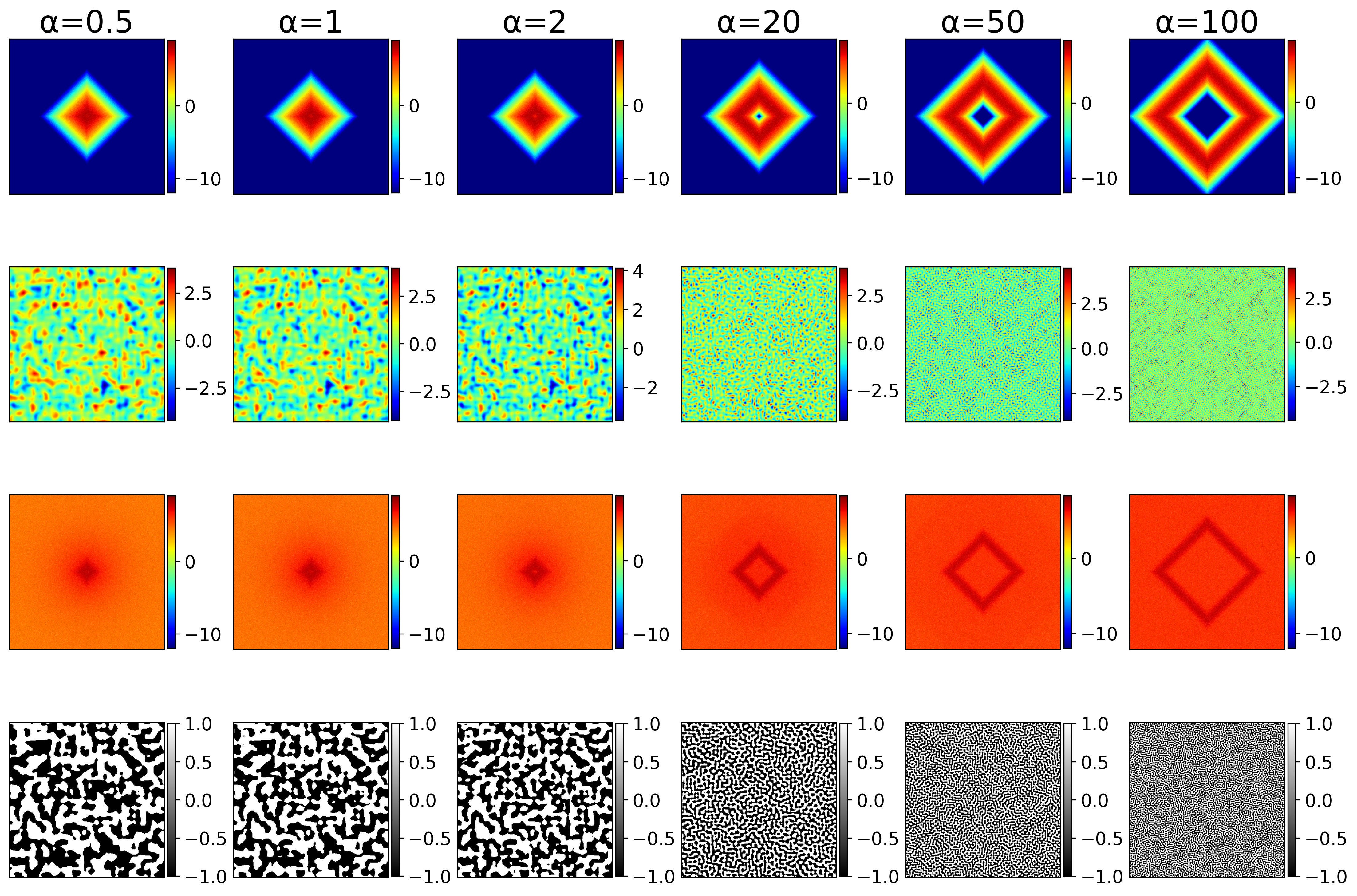}
  \caption{Isotropic case $(a,b)=(1,1)$ with superellipse exponent $p=1$.  Columns correspond to increasing $\alpha$, rows show (from top to bottom): log--spectral density of the continuous field, real--space snapshot of the continuous field, log--spectral density after $\pm1$ thresholding, and snapshot of the binary field.}
  \label{fig:2d_iso}
\end{figure*}

\begin{figure*}[htbp]
  \centering
  \includegraphics[width=0.8\textwidth]{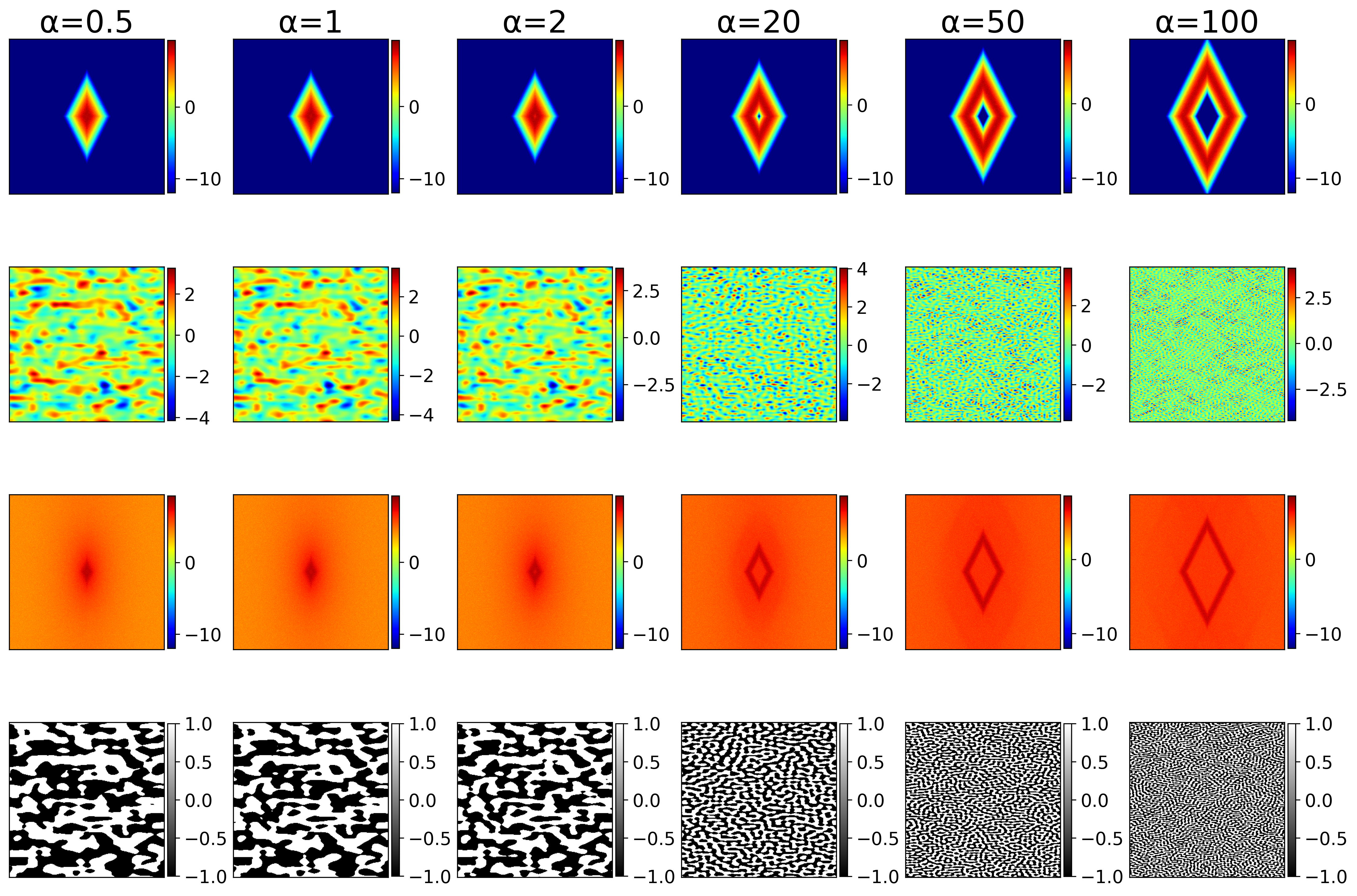}
  \caption{Anisotropic case $(a,b)=(1,0.5)$ with $p=1$.  Layout is identical to Fig.~\ref{fig:2d_iso}.  The horizontal contraction in reciprocal space stretches the spectral shell vertically, introducing directional correlations in real space.}
  \label{fig:2d_aniso}
\end{figure*}

\subsection{Two--dimensional Reconstructions}
Having established both the theoretical underpinnings of hyperuniform spectra and the spectral–filtering algorithm, we now turn to explicit 2D demonstrations.  All simulations are performed on a periodic $1024\times1024$ grid with physical side length $L=100$, giving a lattice spacing $dx=L/N$.  The bandwidth parameter is fixed at $\sigma=2.0$, the target variance is normalised to unity, and the logarithmic exponent is varied over $\alpha\in\{0.5,1,2,20,50,100\}$ to probe the transition from broad to sharply localised spectral shells.  We first consider an isotropic superellipse mask with $(a,b)=(1,1)$ and diamond--like exponent $p=1$, then introduce anisotropy by contracting the minor axis to $(a,b)=(1,0.5)$ while keeping $p$ and all other parameters unchanged.  A single realisation at each parameter set requires approximately 0.03s of compute time, underscoring the efficiency of the proposed pipeline.

Figure~\ref{fig:2d_iso} illustrates how the concentration parameter $\alpha$ governs the radial compactness of the isotropic spectral shell and, consequently, the characteristic length scale of the real–space textures.  When $\alpha\lesssim 1$, the spectral density $\tilde{\chi}_{_\mathcal{K}}(\mathbf k)$ displays a broad diamond‐shaped depletion zone whose half–width is of the same order as its radius, so large–scale fluctuations are only weakly suppressed and the continuous field resembles coloured noise.  As $\alpha$ increases beyond unity, the depletion band contracts into a sharply defined shell and the real‐space morphology coalesces into conspicuous concentric layers that inherit the fourfold symmetry of the mask.  For very large $\alpha$ (e.g.\ $\alpha\ge 20$) the inner hole of $\tilde{\chi}_{_\mathcal{K}}(\mathbf k)$ approaches a genuine exclusion region, mimicking the spectral signature of stealthy hyperuniform systems~\cite{chen2018designing,shi2023computational,ZHONG2025acta}.  In this regime the continuous field is composed of labyrinthine domains with nearly uniform wavelength.  After applying the $\pm 1$ threshold, the resulting binary map still displays a spectral depression that echoes the diamond shell, but the central hole is partially refilled, indicating that strict hyperuniformity is not preserved under the sign transform even though the stealthy–like morphology remains visually evident.

Figure~\ref{fig:2d_aniso} highlights the impact of anisotropy introduced by the aspect‐ratio parameters $(a,b)=(1,0.5)$ while keeping $p$ and $\alpha$ fixed. Compressing the spectral mask along $k_y$ elongates the depletion zone vertically, which in real space translates into stripe‐like domains whose long axis is parallel to the $x$ direction, i.e., perpendicular to the direction of spectral compression. The binary maps reflect the same anisotropic shell morphology, but the spectral density at zero wavenumber remains finite after thresholding, indicating that strict hyperuniform suppression is lost under strong directional bias.

Having established the roles of $\alpha$ and $(a,b)$, we next fix $\alpha$ at a representative hyperuniform value and vary the superellipse exponent $p$ to probe how angular shape alone sculpts both the spectral shell and the resulting continuous and binary morphologies.

\begin{figure*}[htbp]
  \centering
  \includegraphics[width=0.8\textwidth]{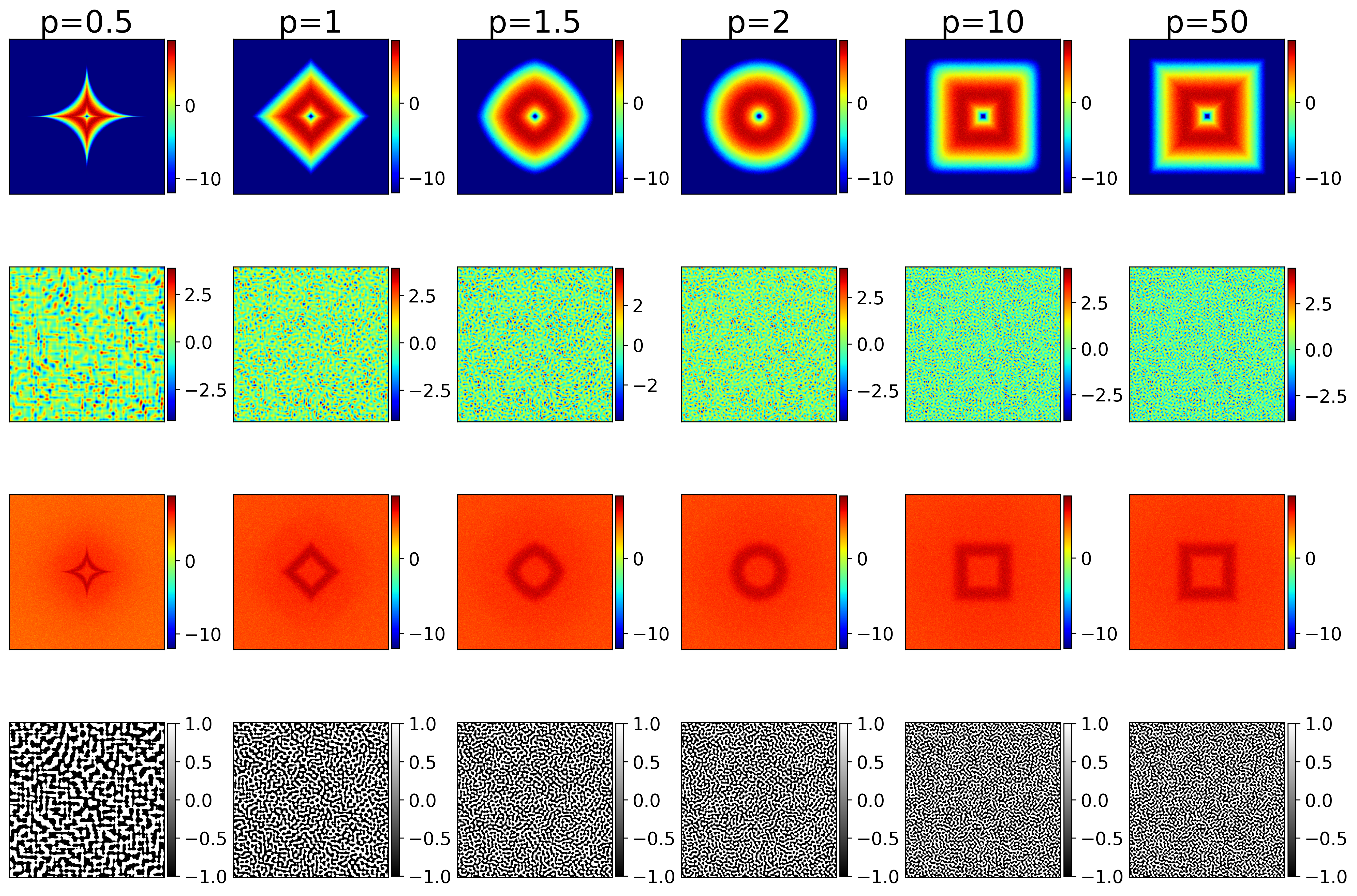}
  \caption{Isotropic masks $(a,b)=(1,1)$ with fixed $\alpha=20$ and six superellipse exponents $p\in\{0.5,1,1.5,2,10,50\}$ (left to right).  Rows, from top to bottom, display the log–spectral density of the continuous field, a $300\times300$ real–space excerpt of the continuous field, the log–spectral density of the $\pm1$ thresholded field, and the corresponding binary excerpt.}
  \label{fig:2d_p_iso}
\end{figure*}

\begin{figure*}[htbp]
  \centering
  \includegraphics[width=0.8\textwidth]{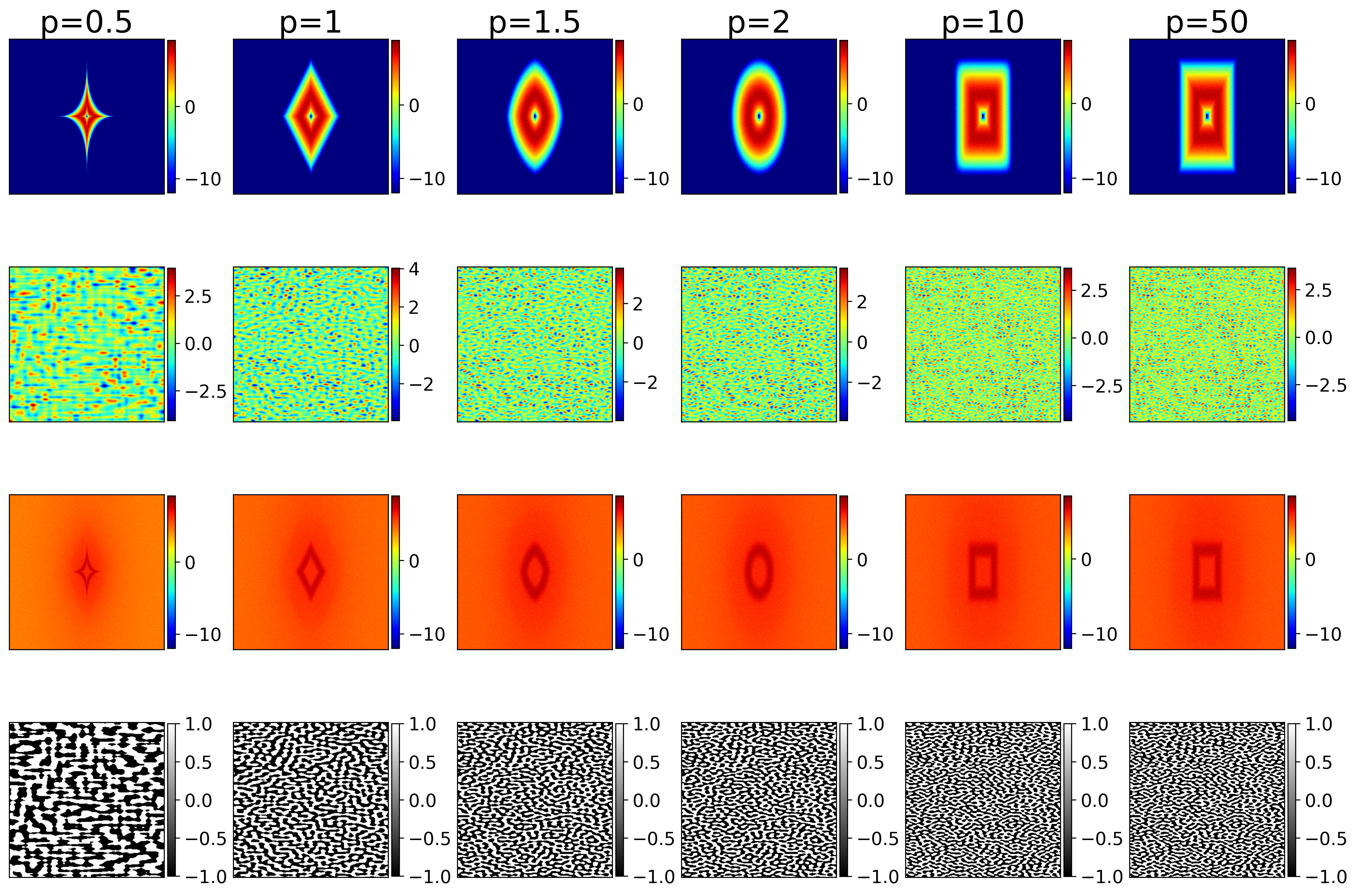}
  \caption{Anisotropic masks $(a,b)=(1,0.5)$ with $\alpha=20$ and the same set of exponents $p$.  Layout parallels Fig.~\ref{fig:2d_p_iso}.}
  \label{fig:2d_p_aniso}
\end{figure*}

Figure~\ref{fig:2d_p_iso} demonstrates the profound influence of the superellipse exponent $p$ on the angular morphology of the isotropic spectral shell at a fixed Class-I hyperuniform concentration $\alpha=20$.  For $p=0.5$ the depletion region assumes a star‐like form with sharp cusps along the Cartesian diagonals, yielding continuous fields composed of radially oriented filaments that intersect to form four–armed junctions.  Increasing $p$ to unity rounds off the cusps, producing a pure diamond shell whose attendant field consists of interconnected rhombic loops.  At $p=1.5$ the shell becomes a squircle, and the real–space labyrinth loses its sharp vertices, acquiring nearly isotropic pore sizes.  When $p=2$ the mask is perfectly circular, giving rise to concentric ring textures devoid of any preferred lattice directions.  Pushing $p$ to $10$ and $50$ progressively squares the shell, and the field reorganises into axis–aligned corridors and orthogonal walls that frame approximately rectangular cavities.  Across all $p$ values the threshold operation blurs fine gradations but preserves the dominant geometric motif: star centres remain visible for $p<1$, rounded annuli persist near $p=2$, and square coronas re-emerge for $p\gg2$.  The zero–wavenumber density rises modestly after thresholding, confirming once more that strict hyperuniformity is sensitive to binarisation, yet the qualitative imprint of the spectral shell survives.

Figure~\ref{fig:2d_p_aniso} repeats the $p$ sweep under a $2{:}1$ aspect ratio, revealing how angular tuning and anisotropy superimpose.  For $p=0.5$ the star-shaped hole elongates vertically, producing continuous fields dominated by chevron bands whose ridges are aligned with the $x$ axis.  At $p=1$ the rhombic shell is likewise stretched, generating diamond loops that appear pinched in the vertical direction.  The squircular shell at $p=1.5$ becomes an oval, and the real–space labyrinth develops elongated pores with gently curved boundaries.  The circular shell of $p=2$ transforms into a clean ellipse, and the field acquires cigar–shaped voids whose long axis is parallel to $x$.  For the square-like masks at $p=10$ and $50$, the corners remain right angles but the opposing sides are unequally spaced, leading to rectangular channels of high contrast.  As in the isotropic case, thresholding tends to soften small-amplitude undulations while leaving the principal geometric skeleton intact.  The gradual evolution from vertically slender stars to horizontally elongated squares underscores the ability of $(p,a/b)$ to decouple radial, angular, and anisotropic control over the hyperuniform shell, furnishing a flexible design space for tailoring pore architecture at a fixed low–wavenumber suppression level.

\subsection{Three--dimensional Reconstructions}

We conclude the Results section by extending the spectral–filtering procedure to volumetric domains.  All 3D realisations are generated on a periodic $256^{3}$ grid with physical side length $L=25$, giving a lattice spacing $dx=L/N$.  The bandwidth and concentration parameters are fixed at $\sigma=2.0$ and $\alpha=20$, respectively, while the superellipse exponent is varied over $p\in\{0.5,1.0,1.5,2.0,10,100\}$.  With these settings a single continuous hyperuniform field is synthesised in $1.2$–$1.3\,\mathrm{s}$, underscoring the method’s scalability from planar to fully three–dimensional grids.

\begin{figure*}[htbp]
  \centering
  \includegraphics[width=1\textwidth]{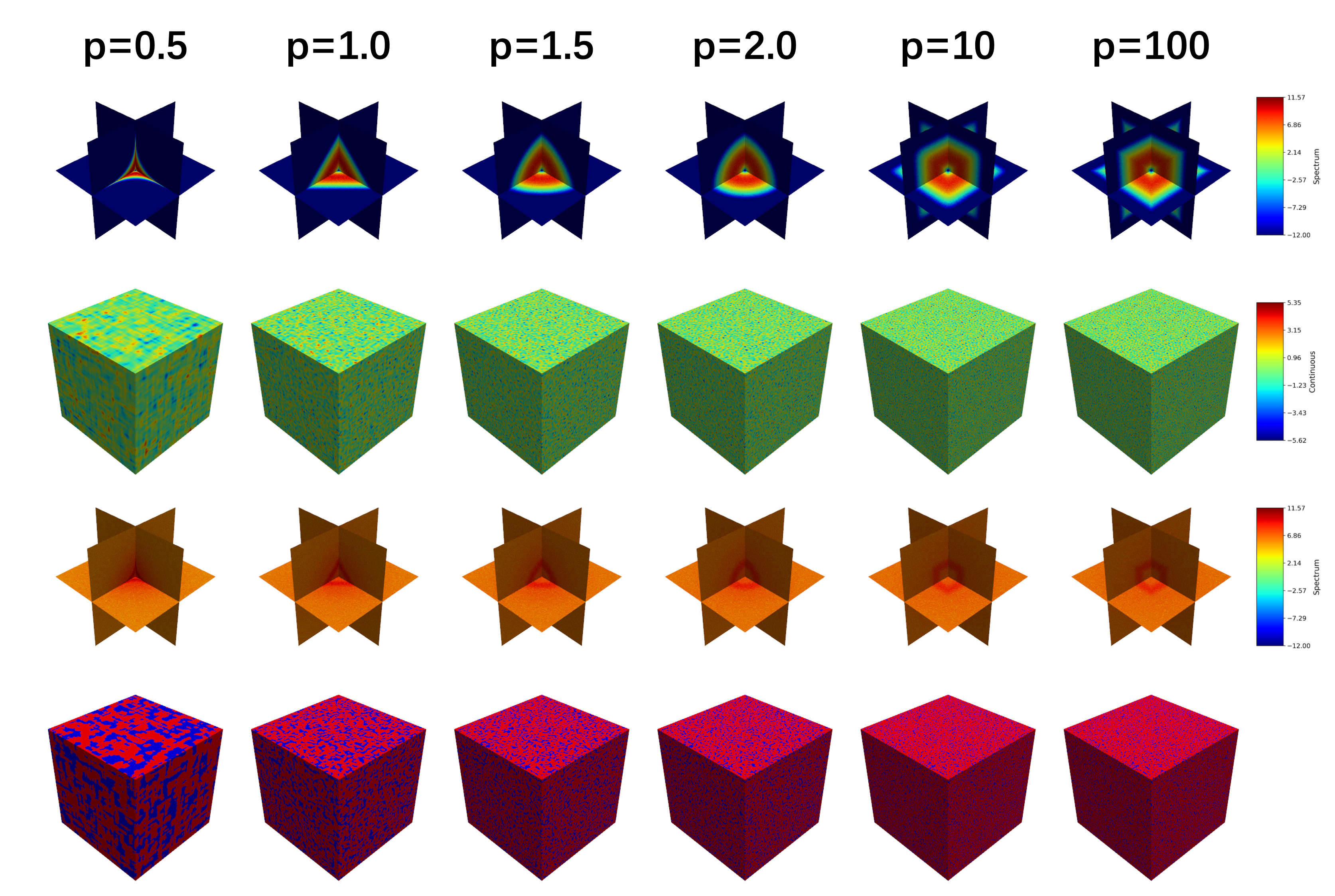}
  \caption{Three–dimensional reconstructions for $\alpha=20$ and six superellipse exponents $p$.  From top to bottom: central $k_{x}\!=\!0$ slice of $\tilde{\chi}_{_\mathcal{K}}(\mathbf k)$, volume‐rendered continuous field, orthogonal planar cuts through the continuous field, and volume‐rendered $\pm1$ thresholded field.  Colour bars refer to the value range of each row.}
  \label{fig:3d_p}
\end{figure*}

Figure~\ref{fig:3d_p} reveals that the qualitative influence of $p$ observed in two dimensions carries over to three.  For $p=0.5$ the spectral density exhibits narrow star–shaped channels extending along the Cartesian diagonals, and the corresponding real–space volume is dominated by thorn–like protrusions, and planar cuts display intricate labyrinths with four–fold cusps.  Increasing $p$ to unity rounds the star into a pure diamond shell, yielding interlaced rhombic corridors that tile the volume.  As $p$ reaches $1.5$ the spectral contour becomes a squircle, and the labyrinth walls lose sharp vertices, giving way to smoothly curved tunnels.  The perfectly circular shell at $p=2$ produces an isotropic array of tube–like pores with a well--defined diameter; the field appears visually homogeneous when rendered at the cube surface.  Pushing $p$ to $10$ and finally $100$ generates near–square shells with crisp right angles; real–space pores align preferentially with the coordinate planes, and the binary volumes are filled by an orthogonal network of rectangular channels reminiscent of a simple cubic lattice.  

Although the $\pm1$ threshold generally reintroduces low–$k$ intensity, the morphological imprint of the underlying spectral shell remains unmistakable: star–shaped perforations for $p<1$, diamond cross–sections near $p=1$, circular or elliptical voids for $p\approx2$, and axis–aligned squares for $p\gg2$.  These observations confirm that the coupled parameters $(p,\alpha)$ govern the geometry of hyperuniform suppression in a dimension–agnostic fashion, providing a powerful handle for tailoring three–dimensional pore architectures at negligible additional computational cost.

\section{Conclusions and discussion}
\label{sec:conclusion}

This study has presented a single shot spectral filtering framework that reconstructs anisotropic hyperuniform continuous random fields with high resolution at a computational cost of only two fast Fourier transforms.  By combining an analytic superellipse mask with the classical spectral representation method, the generator introduces four independent parameters that govern low wavenumber suppression, shell thickness, angular shape and anisotropy.  A $1024^{2}$ realisation is obtained in roughly \(0.03\,\mathrm{s}\) whereas a $256^{3}$ volume is produced in roughly \(1.3\,\mathrm{s}\), which shows that the procedure scales efficiently from planar domains to fully three dimensional grids and greatly reduces the run time compared with iterative reconstruction schemes.

Systematic two dimensional experiments demonstrate clear physical roles for each parameter.  Increasing the concentration exponent contracts the spectral hole, yielding a transition from broad depletion zones to exclusion regions that resemble those of stealthy hyperuniform media.  Changing the aspect ratio stretches the shell, rotates the principal correlation directions and produces stripe like textures in real space without altering the hyperuniform class.  Varying the superellipse exponent sculpts the shell from star to square shapes and imprints the same motifs on both continuous fields and their sign thresholded counterparts.  Three dimensional tests confirm that these trends persist in volumetric settings, where the continuous fields evolve from thorn filled stars to rectilinear cubic networks as the exponent grows.  Although thresholding reintroduces some low wavenumber intensity, the morphological imprint of the shell remains evident, illustrating that the framework can serve as a flexible precursor for designing two phase microstructures with controlled pore architecture.

Future investigations will focus on strategies that preserve exact hyperuniformity after binarisation, the integration of optimisation loops that link the spectral parameters to effective transport and mechanical responses, and the extension of the mask to more elaborate radial or multifractal forms.  The analytic control offered by the parameter quartet provides an accessible route for probing how spectral geometry influences wave propagation, diffusion and conductivity, and it opens a path toward the inverse design of functional materials in photonics, thermal management and multi functional composites.

\noindent{\bf Data Availability Statement}: The codes and data are available upon request.


\smallskip

\bibliography{reference-2}

\end{document}